\documentstyle[aas2pp4,psfig]{article}

\newcommand\mdot   {\hbox {${\dot M}$}}

\newcommand\pers     {s$^{-1}$}

\newcommand\nubreak {$\nu_{\rm b}$}
\newcommand\nuqpo   {$\nu_{\rm QPO}$}

\righthead{The broad-band power spectra of X-ray binaries}
\slugcomment{Submitted to ApJ, Aug 1998}

\begin{document}

\title{The broad-band power spectra of X-ray binaries}

\author{Rudy Wijnands \& Michiel van der Klis}

\affil{Astronomical Institute ``Anton Pannekoek'' and Center for High
Energy Astrophysics, University of Amsterdam, Kruislaan 403, NL-1098
SJ Amsterdam, The Netherlands; rudy@astro.uva.nl,
michiel@astro.uva.nl}

\begin{abstract}
 
We analyzed the rapid aperiodic X-ray variability of different types
of X-ray binaries (black hole candidates, atoll sources, the recently
discovered millisecond X-ray pulsar, and Z sources) at their lowest
inferred mass accretion rates.  At these accretion rates, the power
spectra of all sources are dominated by a strong band-limited noise
component, which follows a power law with an index roughly 1 at high
frequencies and breaks at a frequency between 0.02 and 32 Hz below
which the spectrum is relatively flat.  Superimposed on this, a broad
bump (sometimes a quasi-periodic oscillation) is present with a
0.2--67 Hz centroid frequency that varies in good correlation with the
frequency of the break. The black hole candidates and the
low-luminosity neutron star systems (including the millisecond X-ray
pulsar) have the same relation between the frequency of the bump and
the frequency of the break. These similar characteristics strongly
suggest that in all those different types of sources the band-limited
noise and the bump are produced by the same physical mechanism. This
mechanism cannot then depend on the presence or absence of either a
small magnetosphere or a solid surface, so that it is most likely
related to an instability in the flow in the accretion disk that
modulates the accretion rate.

The Z sources, which are more luminous than the other sources
discussed here, follow a similar, but slightly shifted correlation
between the break frequency and the frequency of the bump. The data
suggest that the band-limited noise in Z sources is more complex than
that in the other sources.

\end{abstract}


\section{Introduction \label{intro}}

At their lowest observed mass accretion rates (\mdot) the
low-luminosity neutron star low-mass X-ray binaries (LMXBs; these are
usually atoll sources; Hasinger \& van der Klis 1989) and the black
hole candidates (BHCs) are remarkably similar with respect to their
X-ray spectra (e.g. Barret \& Vedrenne 1994; see, however, Heindl \&
Smith 1998) and their rapid aperiodic X-ray variability (e.g. van der
Klis et al. 1994a, b; Berger \& van der Klis 1998).  So far, the only
clear observable difference is the detection of quasi-periodic
oscillations (QPOs) between 200 and 1200 Hz in the atoll sources
(Strohmayer et al. 1996; see van der Klis 1998 for a review) and not
in the BHCs (e.g. Remillard et al. 1998).  Apart from these high
frequency phenomena, the power spectra are (almost)
indistinguishable. At their lowest observed \mdot\, the power spectra
of both the atoll sources and the BHCs are dominated by a strong
band-limited noise component, which follows a power law with index of
roughly 1 at high frequencies and breaks at a certain frequency
(\nubreak) below which the spectrum is relatively flat. Superimposed
on this noise component, above \nubreak, a broad bump is present.
This broad bump is sometimes called a QPO although it usually does not
satisfy the commonly used criterium (the ratio of the full width at
half maximum and the centroid frequency is less than 0.5) to be called
QPO.  It was shown that the recently discovered (Wijnands \& van der
Klis 1998a,b) millisecond X-ray pulsar SAX J1808.4--3658 was
indistinguishable with respect to the X-ray spectrum (Gilfanov et
al. 1998; Heindl \& Smith 1998) and the rapid aperiodic X-ray
variability (Wijnands \& van der Klis 1998c) from the atoll sources.
It has been suggested that the same physical mechanism is responsible
for the rapid aperiodic timing behavior in these different types of
sources (van der Klis 1994a,b), and from the properties of the
millisecond X-ray pulsar it was concluded that it most likely
originated outside the magnetosphere (Wijnands \& van der Klis 1998c).

Due to the large flow of data generated by the {\it Rossi X-ray Timing
Explorer} (RXTE) satellite, detailed quantitative comparisons between
atoll and BHC power spectra are only now becoming possible.  In this
Letter, we present a detailed analysis of the low-frequency power
spectra of atoll sources, BHCs, and the millisecond X-ray pulsar. We
show that \nubreak\, is well correlated with the frequency of the bump
for the three different source types, strongly indicating that indeed
the same physical mechanism is responsible in these sources. We also
show that the brightest neutron star LMXBs (the Z sources) at their
lowest observed mass accretion rates have similar power spectra,
however, they follow a somewhat different correlation.

\section{Observations, analysis and results \label{results}}

We used data obtained with the proportional counter array (PCA)
onboard RXTE to study the low-frequency power spectra of the different
types of X-ray binaries. When a strong band-limited noise component
was detected with superimposed on it a broad bump or a QPO (hereafter
usually referred to as QPO), we fitted the 0.004--4096 Hz (or
0.004--2048 Hz when the time resolution was insufficient to go to 4096
Hz) power spectra.  Before fitting the power spectra the dead-time
modified Poisson level (Zhang 1995; Zhang et al. 1995) was subtracted.
The fit function consisted of a broken power law ($P
\propto\nu^{-\alpha_{\rm below, above}}$, where $\alpha_{\rm below}$
and $\alpha_{\rm above}$ are the power law index below and above
\nubreak, respectively) to represent the band-limited noise, and a
Lorentzian with centroid frequency \nuqpo\, representing the QPO.  In
the case of Cyg X-1 and GX 339--4 we used a twice broken power law
instead of a once broken power law to fit the band-limited noise
adequately. The second break occurred at frequencies between 4.6 and
7.3 Hz, well above the frequency of the bump in those sources. Above
this second break the spectrum steepened further with an index of
typically 1.8.  For some atoll sources we had to include an extra
cut-off power law ($P\propto\nu^{-\beta}e^{-\nu/\nu_{\rm cut-off}}$),
representing a noise component at frequencies above 100 Hz (see also
Wijnands \& van der Klis 1998c; Ford \& van der Klis 1998). Sometimes
kHz QPOs were also present, which were fitted with Lorentzians. Errors
on the frequencies were determined using $\Delta\chi^2$ = 1.

\begin{figure}[t]
\begin{center}
\begin{tabular}{c}
\psfig{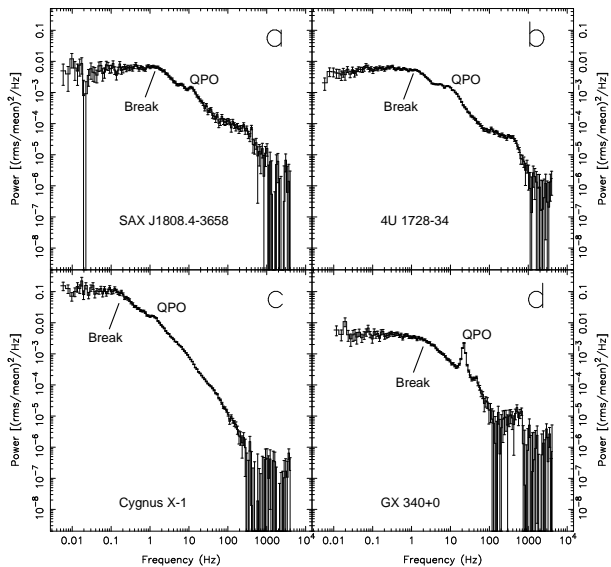}
\end{tabular}
\caption{Typical power spectra of the millisecond X-ray pulsar (SAX
J1808.4--3658; {\it a}), an atoll source (4U 1728--34; {\it b}), a BHC
(Cyg X-1; {\it c}), and a Z source (GX 340+0; {\it d}). The
deadtime-modified Poisson level (Zhang 1995; Zhang et al. 1995) has
been subtracted. \label{powerspectra}}
\end{center}
\end{figure}

In many of the power spectra, one or both of the desired components
were not present.  Usually these power spectra were obtained when the
sources were in different source states compared to those where the
strong band-limited noise and the QPO were detected.  In these states
QPOs are sometimes present, however, the band-limited noise usually is
not. It turned out that the power spectra which were suitable for our
purposes were usually, with a few exceptions (see Section
\ref{thesources}), those taken during the lowest observed inferred
mass accretion rates, i.e., during the low states of the BHCs, during
times when the atoll sources were in their island states, and when the
Z sources were at the leftmost end of their horizontal branches.
Typical power spectra of the source types are shown in
Fig. \ref{powerspectra}.

We also searched the literature for publications of similar power
spectra as observed with RXTE or other X-ray (EXOSAT, Ginga) or
$\gamma$-ray (CGRO/OSSE) satellites. When such power spectra were
encountered we used the values for \nubreak\, and \nuqpo\, as they are
given in those publications.  In Table \ref{data_used} all sources
which were found and their references are listed.  Most sources we
fitted ourselves with the fit function described above. All but two of
the sources found in the literature were fitted using the same fit
function we used; in GRO J0422+52 (Grove et al. 1998) and 1E
1724--3045 (Olive et al. 1998) the band-limited noise was fitted with
a zero-centered Lorentzian and we estimated the break-frequency from
the published power spectra.  Several other sources (e.g. GRS
1737--31: Cui et al. 1997c) also showed the band-limited noise,
however, no evidence for QPOs was found. The very complex sources Cir
X-1 (neutron star system) and GRS 1915+105 (BHC) we left out of our
analysis. Their complex nature, combined with the sometimes many QPOs,
makes it difficult to correctly identify the QPOs and even the break
frequency.

\subsection{The sources \label{thesources}}

We have found eight BHCs (Table \ref{data_used}) which exhibited the
band-limited noise {\it and} the QPO.  We fitted RXTE/PCA data from
the BHCs Cyg X-1, GX 339--4, XTE J1755--324, and GRO J1655--40 (see
also Belloni et al. 1996, Cui et al. 1997a, 1997b, Revnivtsev et
al. 1998, and M\'endez et al. 1998 for more details about the
data). From the literature we obtained the results of the BHCs XTE
J1755--524 (Fox \& Lewin 1998), 1E 1740.7--2942 (Smith et al. 1997),
GRS 1758--258 (Smith et al. 1997), and GRO J0422+32 (Grove et
al. 1998).  Usually the BHCs were in the canonical black hole low
state (GX 339--4; XTE J1755--324; GRO J0422+32; 1E 1740.7--2942; GRS
1758--258; Cyg X-1) but sometimes they may have been in one or two
other previously described states (intermediate or very high; XTE
J1748--288; GRO J1655--40; Cyg X-1; see M\'endez \& van der Klis 1997
for a recent description of BHC source states).  For two sources (1E
1740.7--2942 and GRS 1758--258) a second harmonic to the QPO was
detected and in GRO J1655--40 both a genuine QPO and a broad bump are
present, not harmonically related (see Section \ref{qpo_break}).

We have found eight atoll sources (including the millisecond X-ray
pulsar; Table \ref{data_used}) which exhibited both power spectral
phenomena. Most of the data of these sources we fitted ourselves (see
M\'endez et al. 1997, Wijnands et al. 1998b, and Wijnands \& van der
Klis 1998c for more detailed description of the data), but we also
used results published in the literature (4U 1728--34: Ford \& van der
Klis 1998; 4U 1608--52: Yoshida et al. 1993; 4U 1705--44: Berger \&
van der Klis 1998; Ford, van der Klis \& Kaaret 1998; 1E 1724--3045:
Olive et al. 1998).  All but one of the sources were in the island
state; 4U 1735--44 was in the lower banana branch (Wijnands et
al. 1998b).

Of the Z sources we fitted data of Cyg X-2 (Wijnands et al. 1998a), GX
17+2 (Wijnands et al. 1997), GX 5--1 (Wijnands et al. 1998c), GX 340+0
(Jonker et al. 1998; see also Fig. \ref{powerspectra}d), and Sco X-1
(van der Klis et al. 1996, 1997) when they were at their lowest
observed inferred mass accretion rate (thus at the leftmost end of
their horizontal branches).  The QPO in this case is the fundamental
of the well-known horizontal branch oscillations or HBO. In these
sources the second harmonic to the HBO could frequently also be
detected. We excluded GX 349+2 from our analysis because so far this
source has not exhibited HBO.

\subsection{The QPO frequency versus the break frequency \label{qpo_break}}

In Fig. \ref{break_qpo}a, \nuqpo\, is plotted versus \nubreak\,
obtained for the BHCs (black), the atoll sources (red), and the
millisecond X-ray pulsar (blue). Although intrinsic scatter is
present, \nuqpo\, is well correlated with \nubreak\, for all three
types of sources.  The principal reason for the scatter could be the
complex structure of the QPO. Considerable substructure, usually below
the main peak, is often present (see e.g. Fig. \ref{powerspectra}a and
b). Moreover, in 1E 1740.7--2942 and GRS 1758--258 two QPOs are
detected, harmonically related to each other (Smith et al. 1997). It
is possible that in the other sources also higher or sub harmonics are
present, which are incorporated into the single Lorentzian used to fit
the QPO, resulting in a QPO frequency which is slightly shifted with
respect to the correct value. We did not plot the data of 1E
1740.7--2942 and GRS 1758--258 in Fig. \ref{break_qpo}a (but they are
plotted in Fig. \ref{break_qpo}b) because it is unclear which one of
the two QPO harmonics in those sources is similar to the one fitted in
the other sources. A clue to resolve this might be that the
fundamentals are in both sources only marginally consistent with the
above described relation between \nubreak\, and \nuqpo, while the
second harmonics are completely consistent with it.

Another source for which the identification of the QPO is uncertain is
GRO J1655--40 (also only shown in Fig. \ref{break_qpo}b).  In this
source above \nubreak\, both a genuine QPO near 6.5 Hz and a broad
bump near 9.5 Hz are present (see Fig. 2 of M\'endez et al. 1998). The
frequencies of both phenomena are consistent with the relation between
\nubreak\, and \nuqpo, although the frequency of the genuine QPO
sometimes only marginally so.  Such confusion might also be present in
some of the other sources which do not clearly show both phenomena,
although they could be present.  This again could result in a \nuqpo\,
somewhat shifted with respect to the correct value.

Despite the intrinsic scatter, the relation between \nubreak\, and
\nuqpo\, is very remarkable, extending over three decades in frequency
for both axes, incorporating three different source types (BHCs, atoll
sources, and the millisecond X-ray pulsar). The ratio of \nuqpo\, to
\nubreak\, does not remain constant, but it decreases with increasing
frequency.

We investigated whether the Z sources at their lowest observed mass
accretion rates also would fit this relation. The Z sources are
plotted in Fig. \ref{break_qpo}b (red). Although a correlation between
\nubreak\, and \nuqpo\, does exist here too, it is clear that the Z
sources do not follow the same relationship as the other sources.

\section{Discussion \label{discussion}}

We have presented results on the broad band power spectra of different
types of X-ray binaries.  In BHCs, atoll sources, the millisecond
X-ray pulsar, and the Z sources similar power spectral shapes below
approximately 100 Hz are observed when they are at their lowest
observed mass accretion rates.  Fig. \ref{break_qpo} shows that there
is a good correlation between the break frequency of the band-limited
noise component and the frequency of the QPO that is often observed
above this break. The same correlation applies to the BHC, the atoll
sources, and the millisecond X-ray pulsar, a slightly different one to
the Z sources.  Clearly, in Fig. \ref{break_qpo}b the Z sources are
shifted to the left or to above the data obtained for the other source
types.  Perhaps the QPOs observed in the Z sources are the second
harmonic instead of the fundamental, which would shift the data points
of the Z sources exactly on top of those of the atoll sources and the
BHCs. However, this seems to be ruled out by the tight correlation,
covering atoll and Z sources, between these QPOs and the kHz QPOs
(Psaltis, Belloni, \& van der Klis 1998), indicating that the QPOs in
Z sources are the same phenomenon as those observed in the atoll
sources. This means that the \nubreak\, obtained for the Z sources is
not the one obtained for the atoll sources, perhaps due to the fact
that the strong band-limited noise components are of different origin
in the two types of sources.

\subsection{BHCs, atoll sources, and the millisecond X-ray pulsar}
 
The good correlation between \nubreak\, and \nuqpo\, in BHCs and atoll
sources (including the millisecond X-ray pulsar) suggest that the
band-limited noise and the QPO are caused by one and the same physical
mechanism (see also Wijnands \& van der Klis 1998c).  The similarities
between these source types show that the presence or absence of a
solid surface and a magnetosphere do not affect these rapid X-ray
variability components. These components then most likely originate
somewhere in the accretion disk at a distance of at least several tens
of kilometers from the central compact object outside a possible
magnetosphere, which has a radius of approximately 30 km for the
millisecond X-ray pulsar (Wijnands \& van der Klis 1998b). However,
the large amplitudes of the band-limited noise (up to 50\% rms)
exclude that the emission carrying these fluctuations originates this
far out in the accretion disk, because most of the gravitational
energy of the accretion disk is released closer to the compact
object. The apparent lack of inclination effects excludes, from a
statistical point of view, line of sight obscuration as a mechanism.
A modulation of the accretion rate due to instabilities in the flow in
the region of the disk outside several 10 km from the compact object
then remains as the most likely mechanism for generating the
band-limited noise and the QPO.  Another conclusion is that the HBOs
observed in the Z sources, if they are indeed due to the same physical
mechanism as the QPOs seen in the atoll sources (see above), cannot be
explained by magnetospheric beat frequency models (Alphar \& Shaham
1985; Lamb et al. 1985).

The question arise which fundamental properties of the sources
determine the exact values of \nubreak\, and \nuqpo.  The effect of
the magnetic field of the compact object is probably small in the
region of the disk where the frequencies are determined. The other
physical parameters which can affect the accretion disk are the mass
and spin of the central object, and the mass accretion rate. For a
given source the mass and spin do not change and most likely only the
mass accretion rate determines \nubreak\, and \nuqpo\, (both are
thought to be positively correlated with \mdot; e.g. van der Klis
1994b). Usually the relation between the frequencies and the inferred
mass accretion rate is an one-to-one relation, however, it was shown
that this is not the case for the millisecond X-ray pulsar (Wijnands
\& van der Klis 1998c). The reason for this is as yet unknown, but it
could be related to the transient nature of this source (see also
Wijnands \& van der Klis 1998c).  The difference between the sources
with similar mass accretion rates would be the mass or the spin rate,
or both, of the compact object. This could explain why on average the
BHCs have smaller frequencies than the neutron star systems. However,
considerable overlap between these source types occur, indicating that
the mass accretion rate differences dominate the frequencies.

\subsection{The Z source Sco X-1 in more detail}

When examining the power spectra of Sco X-1, an extra noise component
is present in the frequency region between the band-limited noise and
the fundamental of the QPO (van der Klis et al. 1997; see also
Fig. \ref{sco_powerspectrum}). A similar noise component just below
the QPO can also be observed in GX 17+2 (Homan et al. 1998). Perhaps
this extra noise component is similar to the band-limited noise
observed in the other types of sources. The dominant band-limited noise
in the Z sources is then something different, and could be related to,
e.g., the 6--7 Hz QPO seen in Z sources at higher mass accretion rates
(see also van der Klis et al. 1997).

\begin{figure}[t]
\begin{center}
\begin{tabular}{c}
\psfig{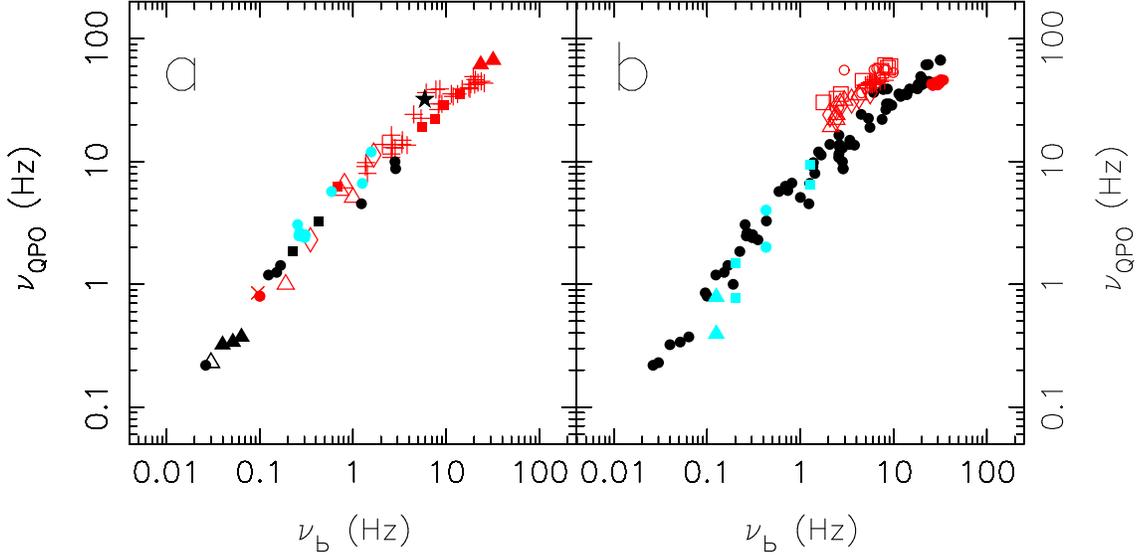}
\end{tabular}
\caption{The frequency of the QPO versus the break frequency of the
band-limited noise. In {\it a} the black points are black hole
candidates (Cyg X-1: filled circles; GX 339--4: filled triangles; XTE
J1748--324: filled star; XTE J1755--324: filled squares; GRO J0422+32:
open triangle), the red points are atoll sources, (4U 1728--34:
plusses; 4U 0614+09: filled squares; 4U 1608--52: open triangles; 4U
1735--44: filled triangles; 4U 1812--12: cross; 4U 1705--44: open
diamonds; 1E 1724--3045: filled circle), and the blue filled circles
are the millisecond X-ray pulsar SAX J1808.4--3658. In {\it b} the
black points are the BHCs, the atoll sources, and the millisecond
pulsar, the red points are the Z sources (Cyg X-2: open circles; GX
17+2: open squares; GX 5--1: open diamonds; GX 340+0: open triangles;
Sco X-1: plusses), and the blue points are the sources (GRO J1655--40:
filled squares; 1E 1740.7--2942: filled circles; GRS 1758--258: filled
triangles) for which two QPOs (sometimes harmonically related) are
present (both frequencies are plotted).  The red filled circles at the
upper right in {\it b} are Sco X-1, however, to determine \nubreak\,
not the band-limited noise was used but the extra noise component at
$\sim30$ Hz (see text). Error bars are smaller than the size of the
data points. 
\label{break_qpo}}
\end{center}
\end{figure}

\begin{figure}[t]
\begin{center}
\begin{tabular}{c}
\psfig{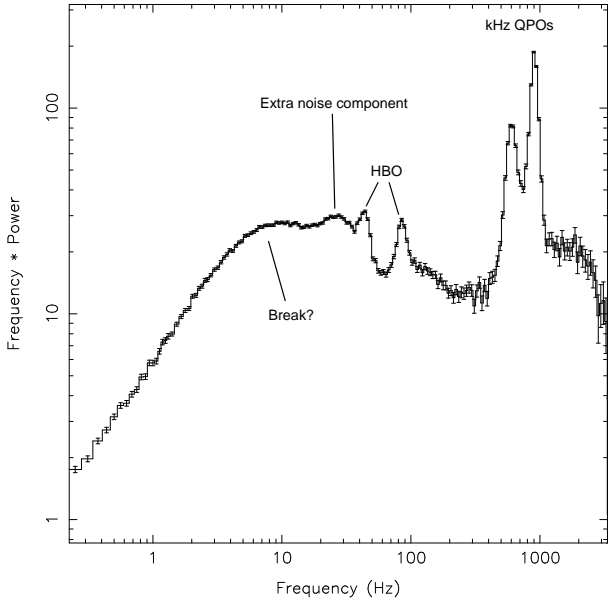}
\end{tabular}
\caption{Typical Leahy-normalized power spectrum of the Z source Sco
X-1 (see also van der Klis et al. 1997), indicating the extra noise
component at $\sim$30 Hz. On the y-axis not the Leahy power but the
frequency times the Leahy power (in units of
[(counts~\pers)(rms/mean)$^2$)]) has been used in order for the
features to be more clearly visible in the power spectrum (see also
Belloni et al. 1997).  Clearly visible are both the lower frequency
and the higher frequency kHz QPO and the fundamental and second
harmonic of the HBO; the extra noise component is the bump below the
HBO fundamental.  The continuum at high frequencies is affected by
instrumental effects.
\label{sco_powerspectrum}}
\end{center}
\end{figure}

In order to test the hypothesis that this extra noise component in Sco
X-1 is similar to the band-limited noise in the other source types, we
fitted the band-limited noise in Sco X-1 with a Lorentzian with a
centroid frequency near zero Hertz and the extra noise component with
a broken power law. The resulting data points (\nubreak\, between 25.5
Hz and 31.9 Hz; \nuqpo\, between 41.6 and 46.4 Hz) are shown in
Fig. \ref{break_qpo}b as the red filled circles. The points are
shifted to higher \nubreak\, and, taking into account the systematic
effects introduced by using a different fit function, are consistent
with the same relation as the other sources. At this point, the
relation of the band-limited noise and the extra noise component in
the Z sources to the band-limited noise in the other sources is not
entirely clear.

\acknowledgments

This work was supported in part by the Netherlands Foundation for
Research in Astronomy (ASTRON) grant 781-76-017. We thank many people,
including several guests, at the Astronomical Institute ``Anton
Pannekoek'' of the University of Amsterdam for their discussions about
the low-frequency power spectra of X-ray binaries. Particularly, we
wish to thank Eric Ford for providing the data on 4U 1728--34 and 4U
1705--44. We thank the anonymous referee for his helpful comments on
the paper.

\clearpage

\clearpage

\newpage

\begin{deluxetable}{lllllll}
\tablecolumns{7}
%\footnotesize
\tiny
\tablewidth{0pt}
\tablecaption{Sources with band-limited noise and QPO\label{data_used}}
\tablehead{
Source  & Type$^{\rm a}$  & Break frequency$^b$ & QPO frequency$^b$ &
Satellite &   References$^c$ \\
        &                 &    (Hz)         &     (Hz)      &  &           &      }
\startdata
Cyg X-1         & BHC             &   0.026--2.9    &  0.29--10.0   & RXTE &
1, 2, 3, 4, 5\\
GX 339--4       & BHC & 0.040; 0.064 & 0.32; 0.37 & RXTE & 1, 2\\
XTE J1748--288  & BHC & 5.9 & 32 & RXTE & 6\\
XTE J1755--324  & BHC & 0.2--0.4 & 1.9--3.3 & RXTE & 2, 7\\ 
GRO J1655--40   & BHC & 0.20; 1.28 & 0.77 or 1.49; 6.48 or 9.48$^d$ &
RXTE & 2, 8\\
GRO J0422+32    & BHC & 0.03$^e$ & 0.23 & CGRO/OSSE & 9\\
1E 1740.7--2942 & BHC & 0.43     & 2.0$^f$ & RXTE & 10\\
GRS 1758--258   & BHC & 0.125    & 0.394$^f$ & RXTE & 10\\
\hline
4U 1728--34     & A   & 1.4--25.7 & 8.0--49.2 & RXTE & 11\\
4U 0614+09      & A   & 0.69--13.9 & 6.3--35.1 & RXTE & 1, 12\\
4U 1608--52     & A   & 0.2--1.0 & 1.0--6.7  & RXTE; Ginga & 1, 13\\
4U 1735--44     & A   & 23.5; 31.9 & 61.7; 67.0 & RXTE & 1, 14\\
4U 1812--12     & A   & 0.095  & 0.85 & RXTE & 1\\
4U 1705--44     & A   & 0.35; 1.67 & 2.3; 11.3 & EXOSAT; RXTE & 15, 16\\
1E 1724--3045   & A   & 0.1$^e$ & 0.8 & RXTE & 17 \\
\hline
SAX J1808.4--3658 & P & 0.26--1.6 & 2.4--12.0 & RXTE & 2, 18\\
\hline
Cyg X-2 & Z &  3.0--9.9 & 36.1--55.6$^f$ & RXTE & 1, 19\\
GX17+2 & Z &   1.8--9.2 & 30.3--61.1$^f$ & RXTE & 1, 20\\
GX5--1 & Z &   2.1--5.6 & 22.3--37.0$^f$ & RXTE & 1, 21\\
GX 340+0 & Z & 2.1--4.3 & 19.2--34.9$^f$ & RXTE & 1, 22\\
Sco X-1 & Z &  5.0--8.0 & 41.6--46.4$^f$ & RXTE & 1, 23, 24\\ 
\enddata 
\tablenotetext{a}{BHC: black hole candidate; A: atoll source; Z: Z
source; P: millisecond X-ray pulsar}
\tablenotetext{b}{When two data points are present they are given
separately; when more are present only the range of observed
frequencies is given. The archival, proprietary, and public TOO
RXTE/PCA results were obtained by fitting the data (see Section
\ref{results}); the other results were obtained from the
literature. Typical relative errors on the QPO frequencies were
0.1\%--1\% for the Z sources and 1\%--10\% for the other sources;
typical relative errors on the break frequencies were 2\%--20\% for
the all sources, except for Sco X-1 they were typically $\sim$1\%.}
\tablenotetext{c}{1: Archival/proprietary RXTE/PCA data; 2: Public TOO
RXTE/PCA data; 3: Belloni et al. 1996; 4: Cui et al. 1997a; 5: Cui
et al. 1997b; 6: Fox \& Lewin 1998; 7: Revnivtsev, Gilfanov, \&
Churazov 1998; 8: M\'endez et al. 1998; 9: Grove et
al. 1998; 10: Smith et al. 1997; 11: Ford \& van der Klis 1998;
12: M\'endez et al. 1997; 13: Yoshida et al. 1993; 14: Wijnands et
al. 1998b; 15: Berger \& van der Klis 1998; 16: Ford, van der Klis, \&
Kaaret 1998; 17: Olive et al. 1998; 18: Wijnands \& van der Klis 1998c;
19: Wijnands et al. 1998a; 20: Wijnands et al. 1997; 21: Wijnands et
al. 1998c; 22: Jonker et al. 1998; 23: van der Klis et al. 1996; 24:
van der Klis et al. 1997}
\tablenotetext{d}{A QPO and a broad bump present, see text}
\tablenotetext{e}{\nubreak\, estimated from the power spectrum
published in the literature}
\tablenotetext{f}{Presence of a second harmonic}
\end{deluxetable}

\end{document}